\documentclass[10pt,iop]{emulateapj}
\usepackage{apjfonts}
\usepackage{epsfig}
\usepackage{amsmath}
\begin{document}

\title{Multi-Wavelength Afterglows of Fast Radio Bursts}

\author{Shuang-Xi Yi\altaffilmark{1}, He Gao, Bing Zhang}
\affil{Department of Physics and Astronomy, University of Nevada Las Vegas, NV 89154, USA; zhang@physics.unlv.edu}
\altaffiltext{1}{School of Astronomy and Space Science, Nanjing University, Nanjing 210093}

\begin{abstract}
The physical nature of fast radio bursts (FRBs) is not identified. Detecting electromagnetic counterparts in
other wavelengths is essential to measure their distances and to
settle down their physical nature. Assuming that at least some of them are of a cosmological origin,
we calculate their afterglow lightcurves in multi-wavelengths (X-rays, optical and radio) by assuming a range
of their total kinetic energies and redshifts. We focus on forward shock emission, but also consider
the possibility that some of them might have bright reverse shock emission.
In general, the FRB afterglows are
too faint to be detected by current detectors. Only if an FRB has a very low radiative efficiency in
radio (hence, a very large kinetic energy),
and when it is close enough, can its afterglow be detected in the optical and radio bands.
We discuss observational strategies to detect these faint afterglows using future telescopes such as
LSST and EVLA.
\end{abstract}


\section{Introduction}

Fast radio bursts (FRBs) are mysterious transients discovered recently
(Lorimer et al. 2007; Thornton et al. 2013). Their physical origin is
subject to intense debate (e.g. Falcke \& Rezzolla 2014;
Totani 2013; Kashiyama et al. 2013; Popov \& Postnov 2013; Zhang 2014;
Loeb et al. 2014; Kulkarni et al. 2014).
If at least some FRBs are of a cosmological origin,
as indicated by the anomolously large dispersion measure (DM), their redshift
information together with the measured DM offer a powerful tool to study
cosmology, including inferring the baryon content and reionization history
of the universe (Deng \& Zhang 2014; Kulkarni et al. 2014), and directly
constraining cosmological parameters and dark matter equation of state
(Gao et al. 2014; Zhou et al. 2014).

The error boxes of FRBs detected by Parkes multi-beam survey are typically
hundreds of square arc-minutes (Thornton et al. 2013). It is therefore
difficult to pin down their host galaxies and derive their redshifts.
Detecting counterparts of FRBs in other wavelengths would be essential
to localize FRBs. Kashiyama et al. (2013) suggested
binary white dwarf (WD) mergers as the source of FRBs, and proposed possible
associations of some FRBs with Type Ia SNe or X-ray debris disk emission.
Motivated by Swift data
showing evidence of a supra-massive neutron star collapsing into a
black hole (Troja et al. 2007; Lyons et al. 2010; Rowlinson et al. 2010,
2013; L\"u \& Zhang 2014; Yi et al. 2014),
Zhang (2014) suggested possible associations of a small
fraction of FRBs with GRBs. Two tentative associations of FRB-like events
with GRBs may have been
discovered by Bannister et al. (2012)\footnote{A negative search result
was reported by Palaniswamy et al. (2014), but the time windows of some
of these GRBs did not cover the end of the plateau, which is the expected
epoch of FRB emission (Zhang 2014). }. Unfortunately, the redshifts of
the two GRBs were not measured.

Another possibility is to search for the afterglow of FRBs. Zhang (2014)
estimated the brightness of FRB afterglows, and found that it is very
faint owing to their low energetics. He suggested that for a typical
FRB at a cosmological distance, the peak radio afterglow flux is dimmer
than the FRB itself by 6-7 orders of magnitude (at the $\mu Jy$ level).
In this paper, we calculate the multi-wavelength FRB afterglows in detail.

\section{The model}

We apply the standard external shock synchrotron emission afterglow model of GRBs
(M\'esz\'aros \& Rees 1997; Sari et al. 1998; see Gao et al. 2013a for a recent,
detailed review). The simplest afterglow model has several free parameters:
the total kinetic energy $E$,
the initial Lorentz factor $\eta$,
the number density of the ambient medium $n_0$;
the equipartition parameters $\varepsilon_e$, $\varepsilon_B$
for electrons and magnetic fields, respectively; and the electron injection spectral
index $p$. If one considers a pair of (forward and reverse) shocks, the
micro-physics parameters can be different for the two shocks. So altogether one
has nine parameters.

The forward shock (FS) emission component is guaranteed. Whether a
bright reverse shock (RS) emission component exists depends on the unknown
magnetization parameter (the ratio between Poynting flux and matter flux,
usually denoted as $\sigma$) of the outflow (Zhang \& Kobayashi 2005;
Mimica et al. 2009; Mizuno et al. 2009).
Most FRB models invoke highly magnetized neutron
stars or black holes (e.g. Falcke \& Rezzola 2014; Totani 2013; Zhang 2014;
Popov \& Postnov 2013). For example, in the ``magnetic hair'' ejection
model invoking the implosion of a supra-massive neutron star (Falcke \&
Rezzola 2014; Zhang 2014), an FRB is emitted in the ejected magnetosphere.
The outflow is therefore likely highly magnetized at the central engine.
The outflow is accelerated via a magnetic pressure
gradient (e.g. Komissarov et al. 2009; Granot et al. 2011), so that
$\sigma$ decreases with radius with the expense of increasing $\Gamma$.
Significant magnetic dissipation would also occur during the
FRB emission phase. Therefore the $\sigma$ value after the dissipation,
especially at the deceleration radius, is not known. If it is already below
unity, as envisaged in some models (e.g. Zhang \& Yan 2011),
a bright reverse shock emission component may be expected
(Zhang et al. 2003; Zhang \& Kobayashi 2005).

In the following we neglect these complications, and only consider a standard
fireball defined by the total energy $E$ and initial Lorentz factor $\eta$.
For faint afterglows of FRBs, to the first order the
details of jet composision would not affect the global picture.

The deceleration time scale $t_\times$, which is
also the time when the reverse shock crosses the shell (for a non-magnetized
outflow), can be approximated as
\begin{equation}
{t_\times} \sim \frac {l (1 + z)}{2c \eta^{8/3}},
\end{equation}
where $l = {({{3E} \mathord{\left/{\vphantom {{3E} {4\pi {n_0}{m_p}{c^2}}}} \right.
 \kern-\nulldelimiterspace} {4\pi {n_0}{m_p}{c^2}}})^{1/3}}$ is the Sedov length.

Both $E$ and $\eta$ are poorly constrained. The observed FRBs have an energy $E_{\rm FRB}
\sim 10^{38} - 10^{40}$ erg assuming a redshift $z \sim (0.5-1)$ (Thornton et al. 2013).
Observations of
radio pulsars suggest that their radio emission efficiency is typically low,
especially for more energetic ones (Szary et al. 2014). As a result, the total
kinetic energy in an FRB outflow can be significantly greater than the FRB energy.
Within the supra-massive neutron star implosion scenario, the total energy in the
 ejecta is essentially the total magnetic energy of the neutron
star magnetosphere, which can be as large as $\sim 10^{47}$ erg for a magnetar
(Zhang 2014). In the following, we allow $E$ to be in a wide range from $10^{43}-10^{47}$
erg.

Various constraints on
the FRB emission mechanisms suggest that the bulk motion Lorentz factor of an FRB
is at least 100 (e.g. Falcke \& Rezzolla 2014; Katz 2014). In the following,
we adopt a conservative value $\eta = 100$. At $t \gg t_\times$,  the predictions
of afterglow flux do not depend on $\eta$. For a higher $\eta$, $t_\times$
would move to an earlier epoch, and the peak afterglow flux would be increased
accordingly. For $E = 10^{47}$ erg, $\eta=100$, and $n_0 \sim 1~{\rm cm}^{-3}$,
one has $t_\times \sim 3$ s.

The synchrotron radiation spectrum from the FS or RS can be characterized by
a multi-segment broken power law
separated by three characteristic frequencies: the minimum synchrotron frequency
(corresponding to electrons with the minimum Lorentz factor),
the cooling frequency $\nu_c$, and the self-absorption frequency $\nu_a$
(Sari et al. 1998). The peak flux of the spectrum is denoted as $F_{\rm \nu,max}$.
Based on the standard prescription (e.g. Sari et al. 1998; Wu et al. 2003; Yi et al.
2013; Gao et al. 2013a), one can calculate the afterglow emission from FRBs.
At the shock crossing time $t_\times$, the FS emission can be
characterized by
\begin{equation}
{\nu^{f}_{{\text{m}},\times }} = 4.1 \times {10^{16}}\,\varepsilon _{B, f, -2}^{{1 \mathord{\left/
 {\vphantom {1 2}} \right.
 \kern-\nulldelimiterspace} 2}}\,\varepsilon _{e, - 1}^2\,n_0^{{1 \mathord{\left/
 {\vphantom {1 2}} \right.
 \kern-\nulldelimiterspace} 2}}\,\eta _2^4\,{(1 + z)^{ - 1}}\,{\rm Hz},
\end{equation}
\begin{equation}
{\nu^{f}_{c,\times }} = 7.5 \times {10^{19}}\,\varepsilon _{B, f, -2}^{{{ - 3} \mathord{\left/
 {\vphantom {{ - 3} 2}} \right.
 \kern-\nulldelimiterspace} 2}}\,\,n_0^{{{ - 5} \mathord{\left/
 {\vphantom {{ - 5} 6}} \right.
 \kern-\nulldelimiterspace} 6}}\,\eta _2^{{4 \mathord{\left/
 {\vphantom {4 3}} \right.
 \kern-\nulldelimiterspace} 3}}\,E_{47}^{{{ - 2} \mathord{\left/
 {\vphantom {{ - 2} 3}} \right.
 \kern-\nulldelimiterspace} 3}}{(1 + z)^{ - 1}}\,{\rm Hz},
\end{equation}
\begin{equation}
{\nu^{f}_{a,\times }} = 7.4 \times {10^8}\,\varepsilon _{B, f, -2}^{{1 \mathord{\left/
 {\vphantom {1 5}} \right.
 \kern-\nulldelimiterspace} 5}}\,\varepsilon _{e, - 1}^{ - 1}\,n_0^{{3 \mathord{\left/
 {\vphantom {3 5}} \right.
 \kern-\nulldelimiterspace} 5}}\,E_{47}^{{1 \mathord{\left/
 {\vphantom {1 5}} \right.
 \kern-\nulldelimiterspace} 5}}\,{(1 + z)^{ - 1}}\,{\rm Hz},
\end{equation}
\begin{equation}
{F^{f}_{\nu ,\max ,\times }} = 7.8 \times {10^{ - 6}}\,\varepsilon _{B, f, -2}^{{1 \mathord{\left/
 {\vphantom {1 2}} \right.
 \kern-\nulldelimiterspace} 2}}\,n_0^{{1 \mathord{\left/
 {\vphantom {1 2}} \right.
 \kern-\nulldelimiterspace} 2}}\,{E_{47}}\,D_{L,27}^{ - 2}\,(1 + z)\,{\rm Jy}.
\end{equation}
Here the typcial shock micro-physics parameters are normalized to $\varepsilon_e = 0.1$,
$\varepsilon_B = 0.01$, and $p=2.5$.
The evolution of the four parameters  (M\'esz\'aros \& Rees 1997;
Sari et al. 1998; Yi et al. 2013; Gao et al. 2013a)
\begin{equation}
t < t_{\times} :\,\,{\nu^{f}_a} \propto {t^{ \frac{3}{5}}},{\nu^{f}_m} \propto {t^{0}},{\nu^{f}_c}
\propto {t^{-2}},F^{f}_{\nu ,\max } \propto
{t^{3}},
\end{equation}
and
\begin{equation}
t > t_{\times}:\,\,{\nu^{f}_a} \propto {t^{0}},{\nu^{f}_m} \propto {t^{ - \frac{3}{2}}},{\nu^{f}_c}
\propto {t^{-\frac{{1}}{{2}}}},F^{f}_{\nu ,\max }
\propto {t^{0}}.
\end{equation}
Because of a small total energy, an FRB outflow would reach the non-relativistic phase
in a relatively short period of time.
The transition time is when the bulk Lorentz factor
$\gamma-1=1$,
where $\gamma \sim (3E/32 \pi n_0 m_p c^5 t^3)^{1/8}$.
After this transition time, the scaling law of the
FS emission is modified as
 \begin{equation}
{\nu^f_a} \propto {t^{ \frac{6}{5}}},{\nu^f_m} \propto {t^{ - 3}},{\nu^f_c}
\propto {t^{-\frac{{1}}{{5}}}},F^f_{\nu ,\max }
\propto {t^{ \frac{3}{5}}}.
\end{equation}
The non-relativistic phase transition time is roughly $t_{\rm N} \sim
6.6\times 10^4, 1.4\times 10^4, 3.1\times 10^3$ s for
with $E=10^{47}, 10^{45}, 10^{43}$ erg, respectively.

If a putative RS exists, the four parameters $(\nu_m, \nu_c, \nu_a, F_{\rm \nu,max})$ of the two shocks
can be related to each other at $t_\times$, which depend on the {\rm ratios} between the micro-physics
parameters of the two shocks (Kobayashi \& Zhang 2003; Zhang et al. 2003).
When explicitly written, these four parameters are
\begin{equation}
{\nu^{r}_{m,\times}} = 1.3\times 10^{13} \varepsilon _{B,r,-1}^{1/2}\,\varepsilon _{e,-1}^{2} n_{0}^{1/2} \eta_2^{2}{(1 + z)^{ - 1}}\,{\rm  Hz},
\end{equation}
\begin{equation}
{\nu^{r}_{c,\times}} = 2.4\times 10^{18} \varepsilon _{B,r,-1}^{-3/2} n_{0}^{-5/6} \eta_2^{4/3} E_{47}^{-2/3}{(1 + z)^{ - 1}}\,{\rm Hz},
\end{equation}
\begin{equation}
\nu _{a,\times }^{\text{r}} = 7.2 \times {10^{11}}\,\varepsilon _{B,r, - 1}^{{1 \mathord{\left/
 {\vphantom {1 5}} \right.
 \kern-\nulldelimiterspace} 5}}\,\varepsilon _{e, - 1}^{ - 1}\,n_0^{{3 \mathord{\left/
 {\vphantom {3 5}} \right.
 \kern-\nulldelimiterspace} 5}}\,\eta _2^{{8 \mathord{\left/
 {\vphantom {8 5}} \right.
 \kern-\nulldelimiterspace} 5}}\,E_{47}^{{1 \mathord{\left/
 {\vphantom {1 5}} \right.
 \kern-\nulldelimiterspace} 5}} {(1 + z)^{ - 1}}{\rm Hz},
\end{equation}
\begin{equation}
F_{\nu ,\max, \times }^{r} =  2.5\times 10^{-3}\, \varepsilon _{B,r,-1}^{1/2} n_{0}^{1/2} \eta_2 E_{47} D_{L,27}^{-2} (1 + z)\,{\rm Jy}.
\end{equation}
Notice that $\varepsilon_B$ is normalized to 0.1, in view that the outflow is likely magnetized.

The scaling laws of RS before and after the crossing time are (e.g. Kobayashi 2000; Yi et al. 2013; Gao et al. 2013a)
\begin{equation}
t < t_{\times} :\,\,{\nu^{r}_a} \propto {t^{ -\frac{33}{10}}},{\nu^{r}_m} \propto {t^{6}},{\nu^{r}_c}
\propto {t^{-2}},F^{r}_{\nu ,\max } \propto{t^{ \frac{3}{2}}},
\end{equation}
and
\begin{equation}
t > t_{\times}:\,\,{\nu^{r}_a} \propto {t^{ -\frac{102}{175}}},{\nu^{r}_m} \propto {t^{ - \frac{54}{35}}},{\nu^{r}_c}
\propto {t^{-\frac{{54}}{{35}}}},F^{r}_{\nu ,\max }
\propto {t^{-\frac{{34}}{{35}}}}.
\end{equation}

\section{Results}

Figure 1 shows the calculated FS FRB afterglow lightcurves in the X-ray (2 keV, panel a), optical
(R-band, panel b), and radio (1 GHz, panel c) bands, respectively. Three different energies, i.e.
$E=10^{47}$ erg (blue), $10^{45}$ erg (red), and $10^{43}$ erg (green), and three different
redshifts, i.e. $z=0.5$ (dashed), 0.1 (dash-dotted), and 0.01 (solid), have been adopted.
Other parameters are fixed to the typical values: $\eta=100$, $n_0 = 1~{\rm cm}^{-3}$, $p=2.5$,
$\epsilon_{B,f}=0.01$, and $\epsilon_e=0.1$. The sensitivity lines of different detectors in
different energy bands
are also plotted. The black solid line in panel a is the sensitivity
line of {\em Swift}/X-Ray Telescope (XRT), which is $\propto t^{-1}$ early on, and breaks to
$\propto t^{-1/2}$ when $F_{\nu}= 2.0\times10^{-15} {\rm erg\,\,cm^{-2}\,s^{-1}}$ at $t=10^5$ s
(Moretti et al. 2009, D. N. Burrows, 2014, private communication). 
The black solid line in panel b is
the sensitivity line of the Large Synoptic Survey Telescope Array (LSST).
In the survey mode, LSST reaches 24.5 mag in 30 seconds (R. Margutti, 2014, private
communication).
The black solid line in panel c is the sensitivity line of the Expanded Very Large Array
(EVLA)\footnote{The Exposure Calculator can be found at https://obs.vla.nrao.edu/ect/},
which scales as
$\propto t^{-1/2}$ for arbitrarily long exposure times.

In general, the broad-band FRB afterglows are all very faint. Only for a large $E$ and a small $z$,
when the predicted afterglow flux becomes of an observational interest. As shown in Figure 1, the X-ray
afterglow becomes detectable by {\em Swift}/XRT only for the most optimistic case calculated,
i.e. $E=10^{47}$ erg, and $z=0.01$ (panel a).
In the optical R-band (panel b),
the peak magnitude of the FS light curves
are about 17, 22, 26, respectively, for $z=0.01, 0.1, 0.5$ and $E=10^{47}$ erg.
The LSST may catch the peak emission only when $z < 0.2$ for
$E=10^{47}$ erg. In the 1 GHz radio band (panel c),
the peak flux density are about $4.4\times10^{-4}$ Jy,  $4.2\times10^{-6}$ Jy, and $1.5\times10^{-7}$ Jy,
respectively, for $z=0.01, 0.1, 0.5$ and $E=10^{47}$ erg. This would be detected by EVLA only when
$z<0.2$ for $E=10^{47}$ erg. 
The peak time shifts to later times with dereasing frequency.
This suggests that follow-up observations are easier in low frequencies. For example, with
$E=10^{47}$ erg, the peak time in 1 GHz is around 1 day.

We also consider the RS emission from FRBs in Figure 2. Fixing other parameters, we allow
$\varepsilon_{B,r}$ to be higher than $\varepsilon_{B,f}$ in view that the outflow is
likely highly magnetized. Defining $R_{\rm B}=(\varepsilon_{B, r}/\varepsilon_{B, f})^{1/2}$
(Zhang et al. 2003), we calculate the cases for
$R_{\rm B}=2$ (purple), $5$ (blue), and $8$ (red).
We fix $E=10^{47}$ erg and consider $z=0.01$ and $z=0.1$, with the FS emission (green) plotted
as a reference.
One can see that with a large $R_{\rm B}$, the RS component would outshine the FS component,
especially in the optical and radio bands, making it easier to detect.
The afterglow is detectable by LSST and EVLA at $z<0.2$.

To better display how peak time and peak flux depend on $E$ and $z$, in Figure 3 we show the contours
of peak time and peak flux in the $E-z$ space. The three panels are for X-rays (panel a), optical
(panel b) and radio (panel c), respectively.
The X-ray peak time is simply the deceleration time $t_\times$.
For the optical band, the peak time is defined when $\nu_m$ crosses the band.
For the radio band, the peak time is defined when ${\rm max}(\nu_m,\nu_a)$ crosses the band.
For the peak flux, we present two (FS vs. RS) values, with the RS value
presented in the parenthesis (noticing the same $E$- and $z$-dependences of $F^f_{\rm \nu,max}$
and $F^r_{\rm \nu,max}$. Since the sensitivity of LSST in the survey mode is a constant
(24.5 mag), we also plotted two thick lines (24.5 mag) above which
LSST can detect the FS (magenta) and RS (green) emissions, respectively.

Assuming that most observed FRBs are at $z\sim 1$, one can derive the event rate for FRBs
below a certain redshift. Assuming that the total event rate density is a constant, i.e.
$\rho \sim 10^{-3}~{\rm gal}^{-1}~{\rm yr}^{-1}$ (Thornton et al. 2013),
a smaller redshift corresponds to a small volume, and hence, a small event rate.
Taking an event rate of $\sim 10^4 ~{\rm sky}^{-1} ~{\rm day}^{-1}$ at $z \sim 1$
(Thornton et al. 2013), one can draw the expected event rate as a function of $z$ based on
volume correction. This is shown as blue dotted lines in the three contour plots.
Notice that the event rate is subject to large uncertainties. For example,
recently Petroff et al (2014) reported a lack of FRBs at intermediate Galactic latitudes,
which suggests either a possible anisotropy of FRB distribution or a lower event rate.
Our event rate curve is still relevant as long as one re-normalize the $z=1$ event rate
to whatever value determined by future observations.

\section{Summary and discussion}

In this paper, we apply the standard GRB afterglow model to predict possible afterglow emission
from FRBs. We calculate their afterglow lightcurves in X-rays, optical, and radio
by assuming a range of their total kinetic energies and redshifts.
In general, owing to their low energetics, the broad-band afterglow emission is predicted to be
very faint, especially for the FS only. Only if the total kinetic energy of FRBs is very large
(radio efficiency very low), and in rare cases when some of them are close enough to earth, could
their FS afterglows become (barely) detectable by the current instruments. It is unclear whether
there is a bright RS component from FRBs. If so (which requires significant de-magnetization
before deceleration), the chance of detecting FRB afterglow in the optical and radio bands
is higher, even though still challenging.

Since data analyses needed to claim the detection of an FRB take significant time, and
since the X-ray afterglow of an FRB peaks early and decays rapidly, follow up observations of FRBs
with X-ray telescopes (e.g. Swift/XRT) would not be fruitful. A better strategy to detect an X-ray
counterpart of an FRB is to apply a wide-field X-ray
telescope (such as Einstein Probe and Lobster), which may catch an X-ray transient associated with
an FRB. However, such a telescope is still being proposed, and it is believed that several
other types X-ray transients, e.g. supernova shock breakouts (Soderberg et al. 2008), jets from
tidal disruption events (Burrows et al. 2011; Bloom et al. 2011), and putative X-ray transients
due to neutron star - neutron star mergers with a millisecond magnetar engine
(Zhang 2013; Gao et al. 2013b; Yu et al. 2013; Metzger \& Piro 2014)\footnote{Double neutron
star mergers can leave behind a supra-massive rapidly spinning neutron star if the mass
of the two neutron stars are small and the neutron star equation of state is hard (Dai et al.
2006; Gao \& Fan 2006).}, would give rise to brighter
X-ray signals than the FRB afterglow. Detecting X-ray transients associated with FRBs is plausible,
but challenging.

In the optical band, the FS peak time is after $t_\times$, while the RS peak time is at $t_\times$.
Again due to the possible delay of data analysis to claim an FRB discovery, follow-up observations
may not be fruitful. One should also appeal to wide-field optical telescopes, such as GWAC (Paul et
al. 2011). The peak flux is however usually too low to be detected by these telescopes, unless the
source is energetic, nearby, and with a bright RS emission component.
The optical afterglow peak emission can be detected by LSST for nearby energetic events in
the survey mode. However, since only 7-10 square degrees are covered in each 30 seconds exposure
(E. Berger, 2014, private communication),
it still takes great chance coincidence to detect the optical afterglow of an FRB with LSST.

In the radio band, the telescope that detects the FRB can continue to collect data. As a result,
no trigger information is needed to ``follow-up'' an FRB. On the other hand, the afterglow is
faint. For a Jy-level FRB, the peak afterglow flux is in the $\mu$Jy level for the FS component
for typical parameters,
and at most one order of magnitude brighter for the RS component. For optimistic cases
(large $E$ and small $z$), the afterglow flux may reach the mJy level, but the detection rate
for these extreme cases is very low. In general, large radio telescopes with high sensitivity
is needed. In principle, one can use a small radio telescope to trigger an FRB and use a large
telescope to follow up. The peak time of radio afterglow is $10^4 - 10^6$ s (hours to days).
This would be a good strategy if the data processing time to claim an FRB detection can be reduced
to within hours. Follow up observations with EVLA would be able to catch the FRB afterglow if the source is
energetic and nearby.

The afterglow emission signal discussed in this paper is {\em generic} to progenitor models.
It is also the {\em minimum} multi-wavelength signals one would expect to be associated with an FRB.
Subject to progenitor models, an FRB may be accompanied by other brighter signals
(e.g. Kashiyama et al. 2013; Zhang 2014;
Niino et al. 2014), which can be used to differentiate among the progenitor models.

\bigskip

We thank Edo Berger, David Burrows, and Raffaella Margutti for helpful discussions on the
instrumental sensitivities of EVLA, LSST, and Swift XRT.

\newpage
\begin{figure*}
\includegraphics[angle=0,scale=0.5]{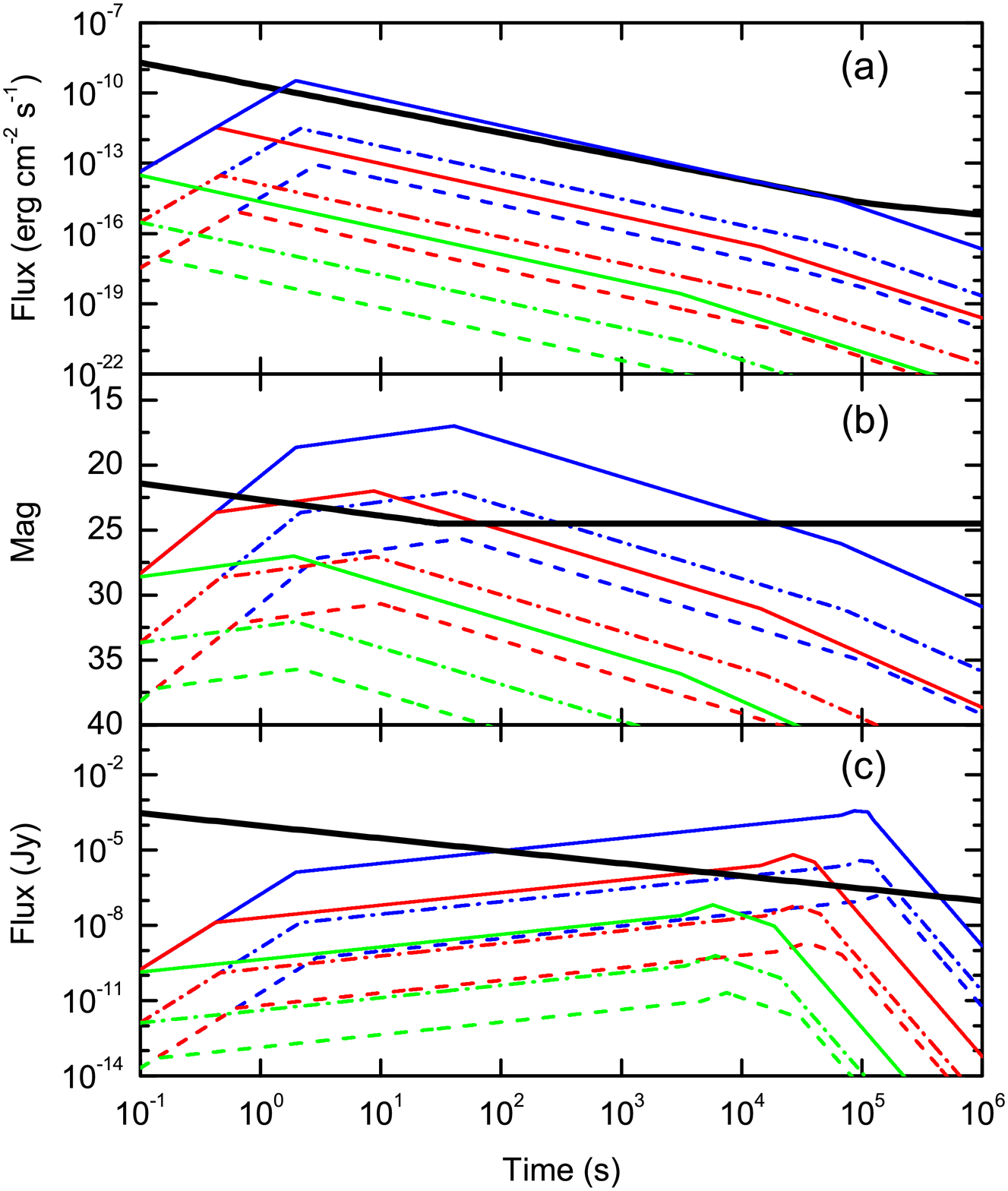}
\caption{Example forward shock afterglow light curves of FRBs. The model parameters: $\epsilon_B=0.01$, $\epsilon_e=0.1$, $n_0=1$, $p=2.5$, and $\eta=100$. Three values of the energy $E=10^{47}$ (blue), $10^{45}$ (red), $10^{43}$ (green), and three values of redshift $z=0.5$ (dashed), $0.1$ (dash-dotted), $0.01$ (solid) have been adopted. (a) The X-ray light curves at 2 keV, the black solid line is the detector sensitivity line of {\em Swift}/XRT; (b) $R-band$ light curves, the black solid line is the detector sensitivity line of LSST; (c) radio light curves at 1 GHz, the black solid line is the detector sensitivity line of EVLA.}
\end{figure*}

\newpage
\begin{figure*}
\includegraphics[angle=0,scale=0.5]{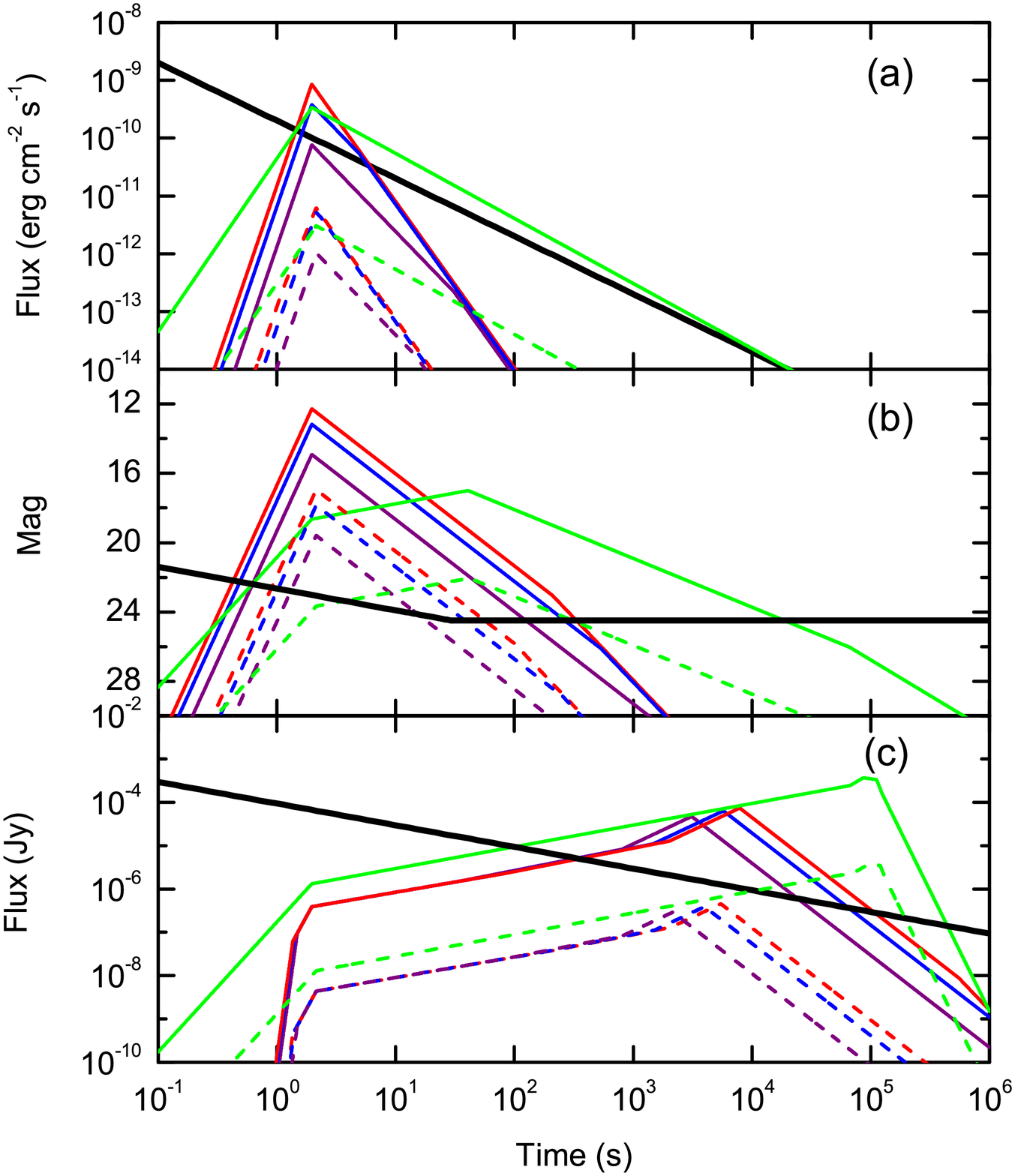}
\caption{Example reverse shock afterglow light curves of FRBs. The model parameters: $\epsilon_e=0.1$, $n_0=1$, $p=2.5$, and $\eta=100$. Only the most optimistic cases with energy $E=10^{47}$ and redshift $z=0.01$ (solid) and $z=0.1$ (dashed) are plotted. Several $R_{\rm B}$ values are adopted to calculate the RS component: $R_{\rm B}=2$ (purple), $5$ (blue), and $8$ (red). The FS component is shown as green in both cases, and the black solid lines are the detector sensitivity lines (same as Fig. 1). (a) The X-ray light curves at 2 keV; (b) $R-band$ light curves; (c) radio light curves at 1 GHz.}
\end{figure*}

\newpage
\begin{figure*}
\centering
\includegraphics[angle=0,scale=0.7,width=0.7\textwidth,height=0.9\textheight]{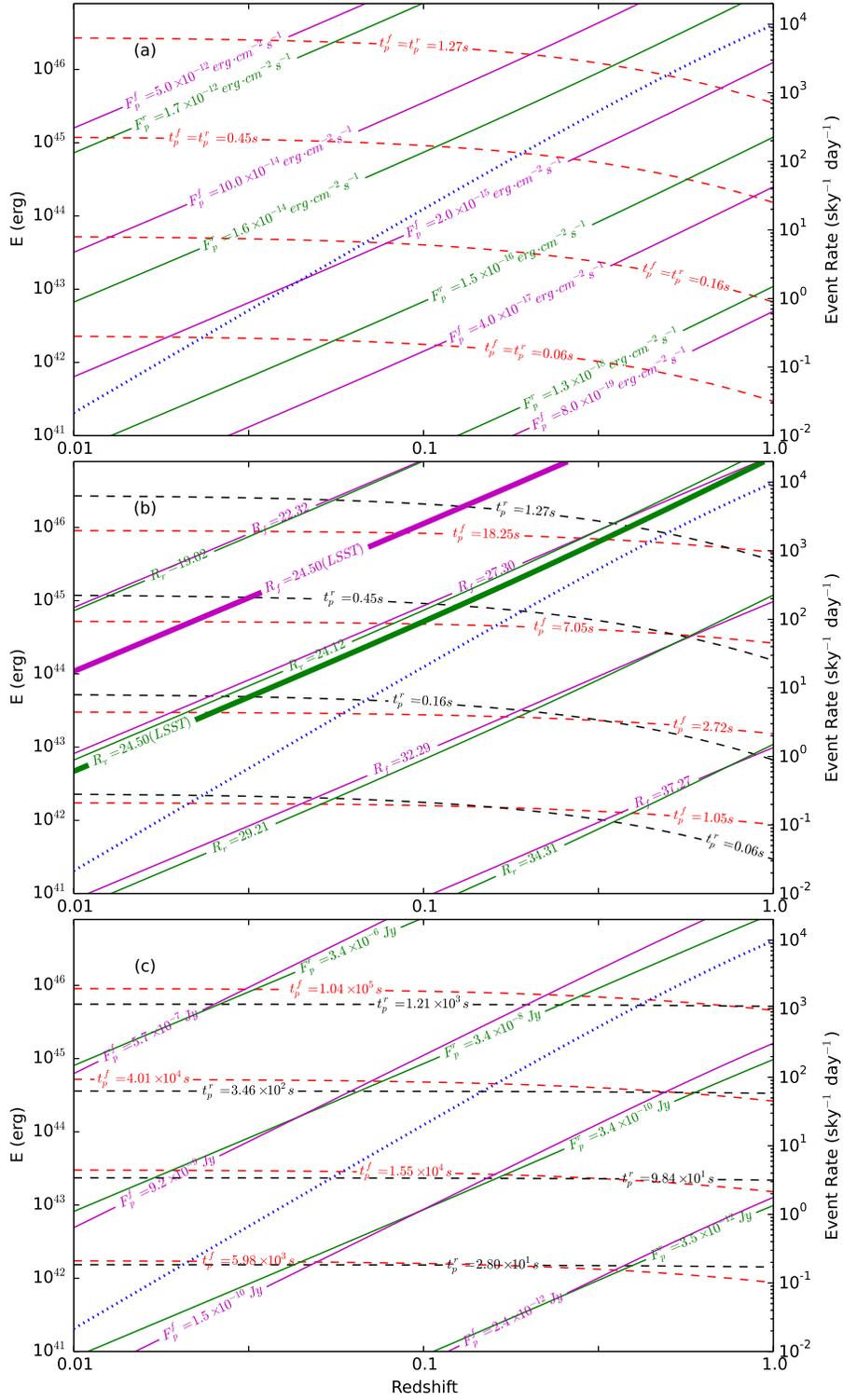}
\caption{Contours of peak time and peak flux in the $E-z$ plane. The peak times are marked in dashed lines (the red for FS and the black for RS), and peak fluxes are marked in solid lines (the purple for FS and the green for RS). The blue dotted line in each panel denote the detection event rate (right vertical label). Panels (a), (b), and (c) are for X-rays, optical, and radio bands, respectively. Thick lines in panel b are the sensitivity lines of LSST in the survey mode:  FS (magenta) and RS (green).}.
\end{figure*}

\end{document}